\documentclass[aip,cha,amsmath,amssymb,reprint]{revtex4-1}
\usepackage{color}
\usepackage{graphicx}% Include figure files
\usepackage{dcolumn}% Align table columns on decimal point
\usepackage{bm}% bold math
\usepackage[mathlines]{lineno}% Enable numbering of text and display math
\usepackage{amsfonts}
\usepackage{amsmath}
\usepackage{sidecap}
\usepackage{float}
\usepackage{parskip}
\usepackage{grffile}

\usepackage{color}
\usepackage{hyperref}
\usepackage{breakurl}
\usepackage{hhline}
\usepackage{mathtools}
\usepackage{comment}
\usepackage[normalem]{ulem}

\makeatletter
\def\@eqnnum{{\normalsize \normalcolor (\theequation)}}
 \makeatother
\begin{document}

%\title{Finite-time Lyapunov signatures of dynamical regimes in reservoir computing}
\title{Finite-time Lyaponov analysis of a trained reservoir computer}

\author{Dishant Sisodia}
\email{dishantsisodia.physics@gmail.com}

\author{Sarika Jalan} \email{sarikajalan9@gmail.com: Corresponding author}
\affiliation{Complex Systems Lab, Department of Physics, Indian Institute of Technology Indore, Khandwa Road, Simrol, Indore-453552, India}

\begin{abstract}

We use finite-time Lyapunov exponent (FTLE) distributions to probe transition mechanisms in high-dimensional reservoir maps trained on low-dimensional chaotic dynamics across multiple regimes. While trained reservoirs accurately predict critical transitions and regime shifts, conventional analyses based on time series or bifurcation structure provide limited mechanistic insight, since distinct pathways in high dimensions can yield similar outputs. We show that FTLE statistics overcome this limitation. This is particularly important for interior crises, where direct identification of unstable periodic orbit collisions in the reservoir space is infeasible. Using the logistic map as a canonical example exhibiting intermittency, fully developed chaos, and crisis-induced transitions, we demonstrate that although such distinct regimes are difficult to characterize within the high dimensional reservoir space, their FTLE distributions are faithfully reproduced. This establishes FTLE analysis as a systematic and reliable framework for uncovering transition mechanisms in learned reservoir dynamics.

\end{abstract}
 
\maketitle

\begin{quotation}
Chaotic systems can undergo abrupt qualitative changes in their behavior. These transitions are sensitively captured by the statistical distributions of finite-time Lyapunov exponents, whose signatures are not accessible through asymptotic Lyapunov exponents and are only coarsely reflected in bifurcation diagrams. Existing studies on reservoir computing primarily rely on such asymptotic indicators, which provide limited insight into the mechanisms driving complex transitions such as interior crises. In this work, we exploit the statistical properties of finite-time Lyapunov exponents to probe what a machine learning model, reservoir computing, learns from time series data. 
The logistic map, a minimal system exhibiting fully developed chaos, intermittency, and crisis-induced transitions, serves as an ideal testbed, with analytical results at the Ulam point enabling direct comparison with the RC model.
We show that FTLE analysis can reveal whether a trained reservoir captures the dynamical changes underlying sudden transitions, even when its high-dimensional internal dynamics are difficult to interpret directly. This demonstrates that FTLE-based analysis is a practical tool for understanding how machine-learning models encode critical transitions and, beyond reservoir computing, may provide a general framework for probing how other architectures learn and represent such behavior.

\end{quotation}

\paragraph{\textbf{Introduction:}}
Lyapunov exponents play a central role in the characterization of dynamical systems, providing quantitative measures of the average rates of separation or contraction of nearby trajectories. A positive Lyapunov exponent is the defining indicator of chaos, signifying exponential divergence of nearby trajectories \cite{Ott2002}. In practical applications, however, asymptotic Lyapunov exponents are often of limited use because experimental and numerical data are necessarily finite in duration. This challenge prompted the study of finite-time Lyapunov exponents (FTLEs), which are computed over finite windows and explicitly depend on initial conditions. Prasad and Ramaswamy \cite{PhysRevE.60.2761} studied the statistical distributions of these exponents in low-dimensional chaotic systems and demonstrated that they exhibit distinct, characteristic functional forms associated with different classes of dynamical behaviors. Such signatures can not be captured by the asymptotic Lyapunov exponent even when there exists no restrictions on size of the available datasets.

Machine learning has become an important tool for studying nonlinear dynamical systems \cite{Tsironis2025,Brunton2016}. Among its approaches, reservoir computing (RC) is particularly effective for temporal data due to its high-dimensional projection \cite{JaegerHaas2004, LuHuntOtt2018, PhysRevResearch.3.013090, PhysRevE.110.034211, shrimali, mandal, PhysRevE.110.034204, 10.1063/5.0252908,10.1063/5.0255707,10.1063/5.0283386}. Beyond prediction, recent work has begun to analyze machine learning algorithms from a dynamical systems perspective. Trained RC architectures have been shown to capture scaling laws of boundary crises \cite{8b7z-fnjd, Kong_2021}, fixed points and periodic orbits \cite{PhysRevE.104.044215}, and to recover topological invariants such as correlation dimension and entropy \cite{PhysRevE.102.033314}. Storm \textit{et al.} \cite{PhysRevLett.132.057301} used Lyapunov exponents to study sensitivity in deep feed-forward networks, and the asymptotic Lyapunov spectrum of high-dimensional RC models has been shown to closely match that of the underlying system \cite{Jpathak}. However, these studies did not examine different dynamical regimes or the mechanisms behind their emergence. In particular, since the trained RC is a high-dimensional system, agreement at the level of projected bifurcation diagrams does not uniquely determine the underlying mechanism of the dynamical transition in the RC, as multiple pathways can lead to similar asymptotic behavior \cite{GREBOGI1983181}. 

In this work, we analyze the distributions of FTLEs of a reservoir map trained on the logistic map at few parameter values, and use them to characterize the learned dynamics across multiple chaotic regimes. We examine the statistical properties of the maximal FTLEs and show that their distributions provide a clear signature of the interior-crisis transition in the reservoir. This is particularly significant because interior crises are caused by collisions with unstable periodic orbits \cite{PhysRevLett.48.1507}, whose direct identification is not possible in the high-dimensional reservoir map. We further establish correspondence between the reservoir and the original system by comparing FTLE distributions across parameter values spanning distinct regimes, including typical chaos, fully developed chaos, and intermittency, demonstrating a systematic dynamical framework for uncovering transition mechanisms in learned high dimensional reservoir systems.

%This study demonstartes that the dynamical changes in the trained RC map, closely mirror those of the low dimensional original system despite the high dimensionality of the reservoir.
\begin{figure}[t!]
    \centering
    \includegraphics[width=0.9\columnwidth]{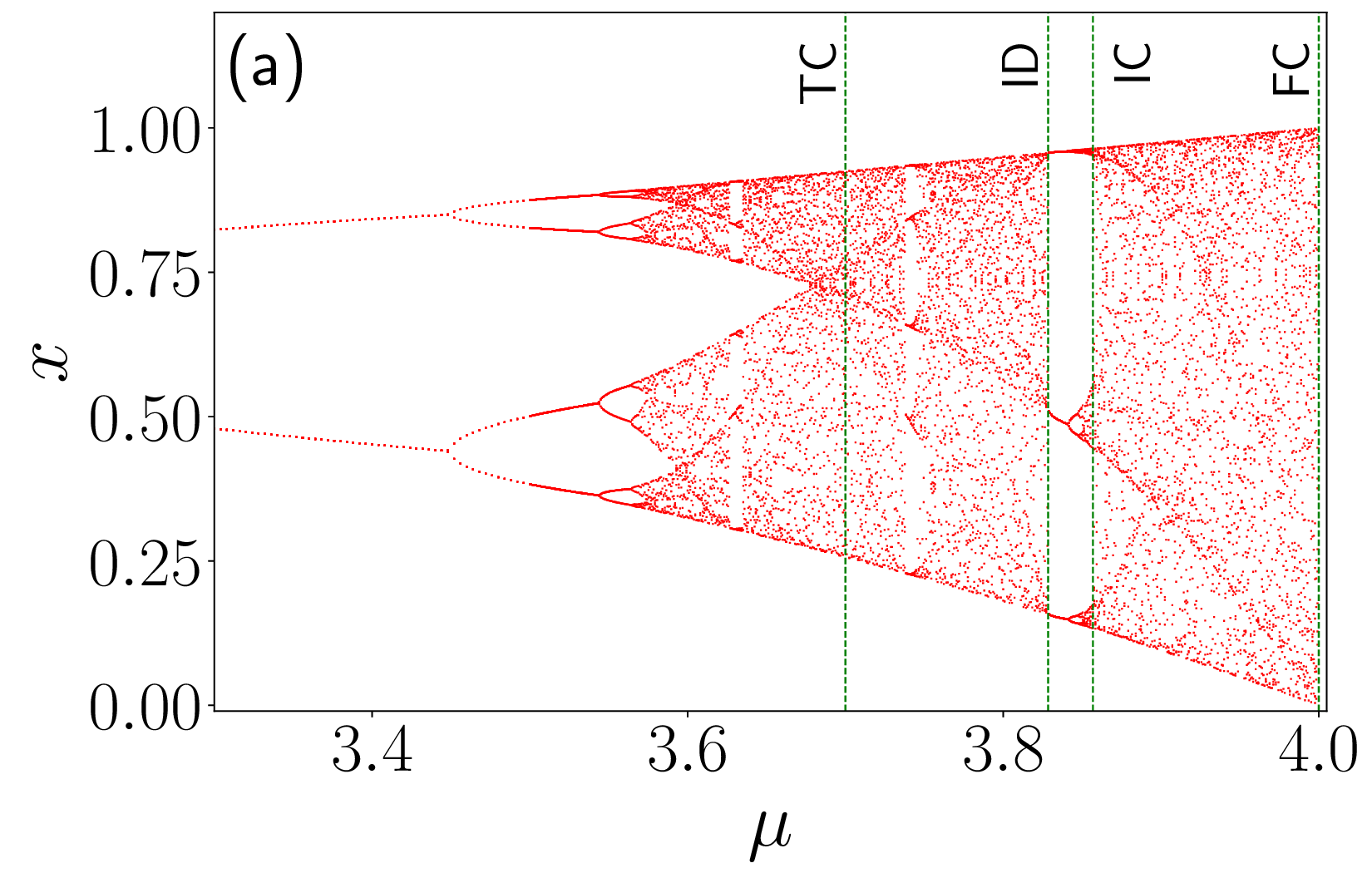}\\

    \includegraphics[width=0.9\columnwidth]{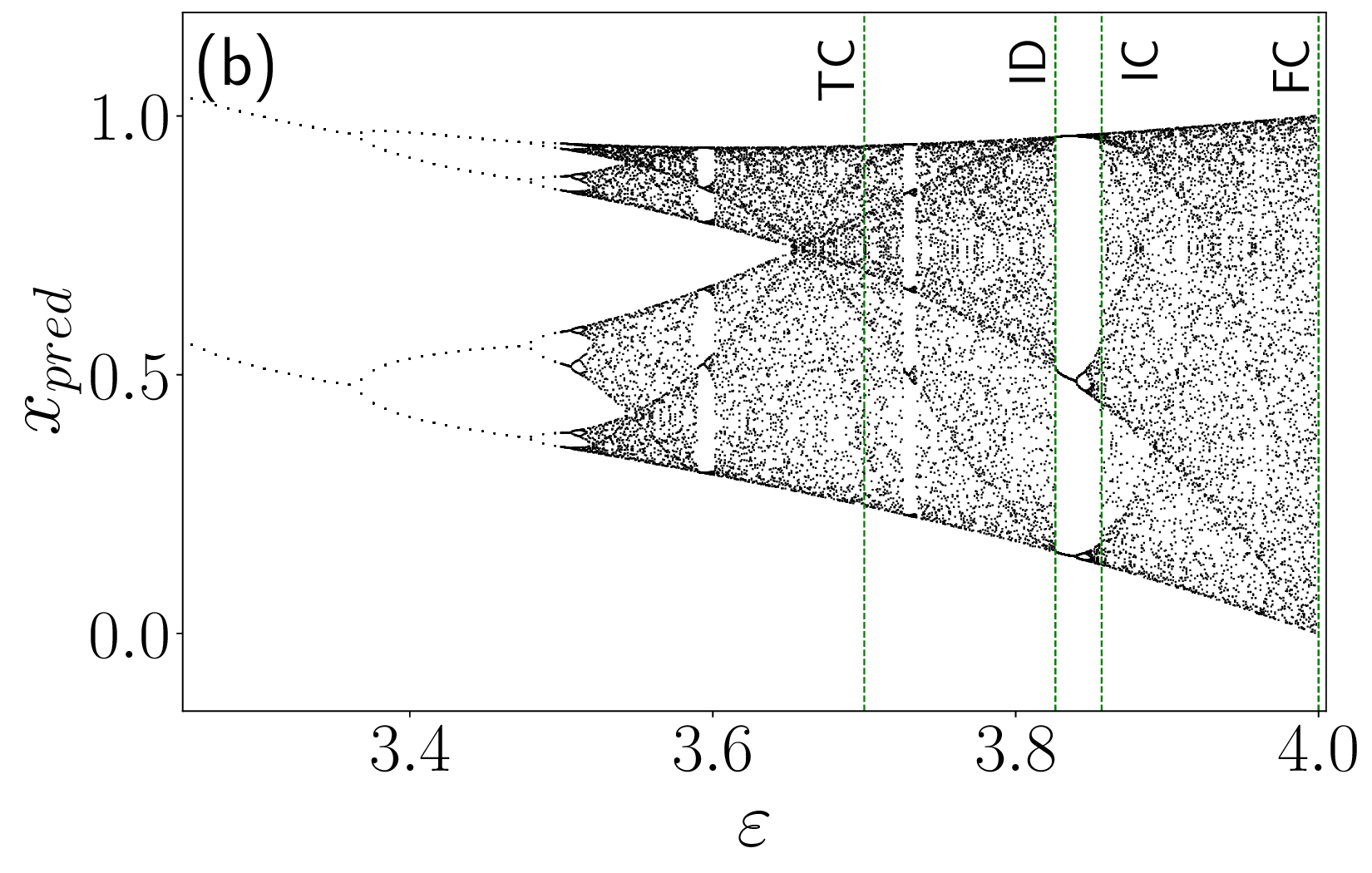}

    \caption{RC mimicking the dynamical regimes of logistic map:
    Bifurcation plot of (a) logistic map and (b) trained RC map (Eq. \ref{eq2}). TC, ID, IC, FC denote typical chaos, intermittent dynamics, interior crisis, fully developed chaos, respectively.
    }
    \label{fig1}
\end{figure}

\paragraph{\textbf{RC Model and FTLE computation:}}
For a dynamical system of $D$ dimensions, the $D$ Lyapunov exponents of an orbit measure the rate at which volume elements in the phase space expand or contract along the orbit, in different directions. The largest asymptotic Lyapunov exponent is defined as \cite{Ott2002}
\begin{equation}\label{eq0}
    \lambda_m = \lim_{t \to \infty} \frac{1}{t} ln \frac{d(t)}{d(0)}, 
\end{equation}
where $d(0)$ is the infinitesimal initial separation between two trajectories and $d(t)$ is their separation after time $t$. The exponent obtained is an asymptotic ($t \to \infty$) quantity. The trajectory is divided into segments of length $N$, to compute the largest FTLE \cite{PhysRevE.60.2761}. For each $k^{th}$ segment:
\begin{equation}\label{eq1}
    \lambda_N (k) = \frac{1}{N} \sum_{j = N(k-1) + 1}^{kN} \lambda_1 (j),
\end{equation}
where $\lambda_1 (j)$ is the single step maximum Lyapunov exponent. To compute $\lambda_N$, we initialize the system with a reference trajectory and an infinitesimally perturbed copy. Both the trajectory and the perturbation are evolved simultaneously, and the instantaneous (single-step) stretching rates $(\lambda_1)$ are recorded at each iteration over a long time series. The maximal $N$-step FTLE is then obtained by applying a moving-window average of size $N$ to these single-step exponents (Eq. \ref{eq1}). Consecutive windows overlap by $N-1$ steps, yielding an ensemble of $\lambda_N$ values. The FTLEs depend sensitively on initial conditions. Denoting by $P(\lambda, N)\, d\lambda$ the probability that the $N$-step FTLE $\lambda_N$ lies in the interval $[\lambda, \lambda + d\lambda]$, it has been shown that the resulting distribution acquires a characteristic shape that reflects the underlying dynamical state of the chaotic attractor~\cite{PhysRevE.60.2761}. For sufficiently large $N$, these local exponents converge to the asymptotic exponent ($\lim_{N \to \infty} P(\lambda, N) \to \delta(\Lambda - \lambda)$).
\begin{figure}[t!]
    \centering
    \includegraphics[width=9cm,height=5cm]{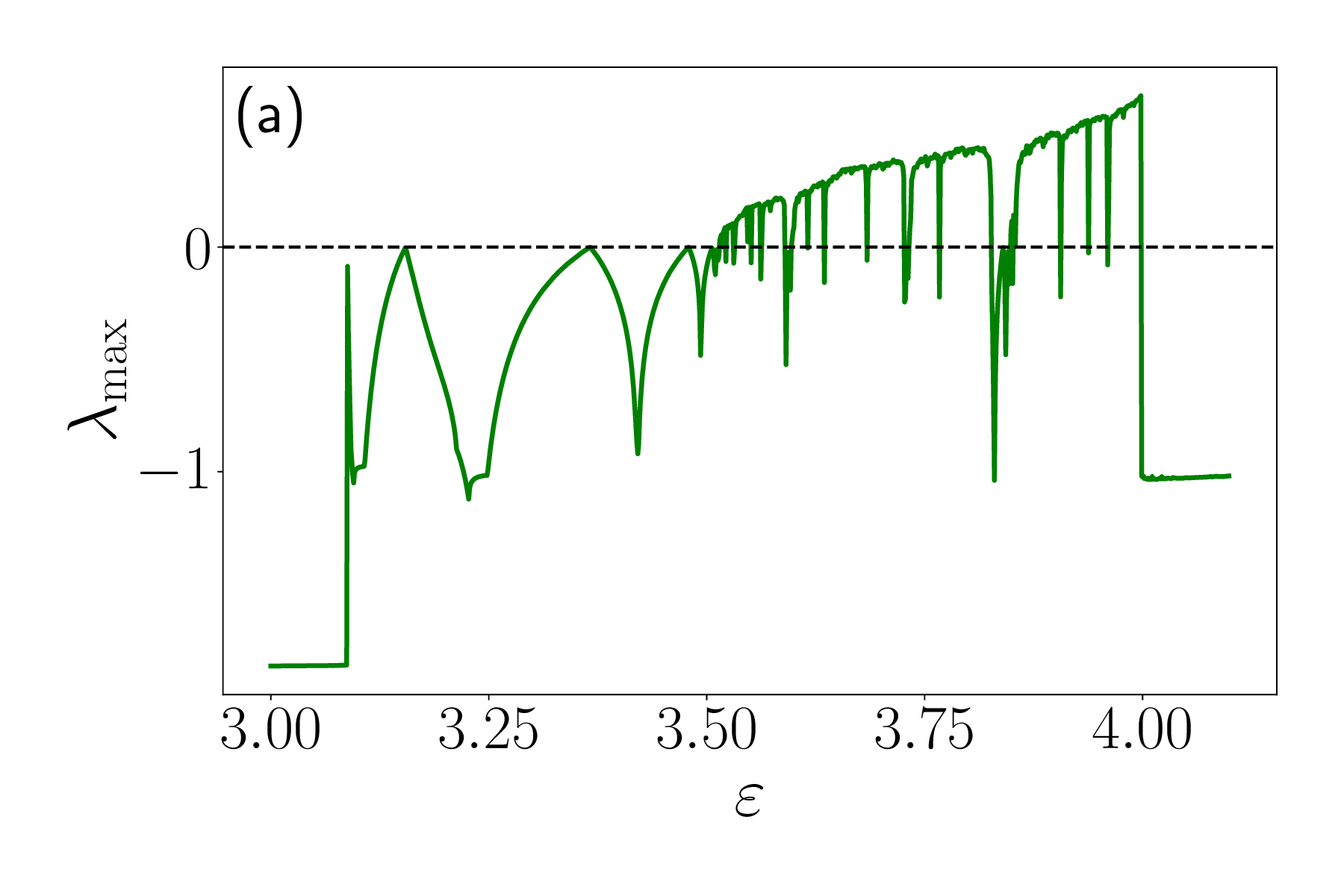}\\
    \vspace{-4mm}
    
    \includegraphics[width=9cm,height=5cm]{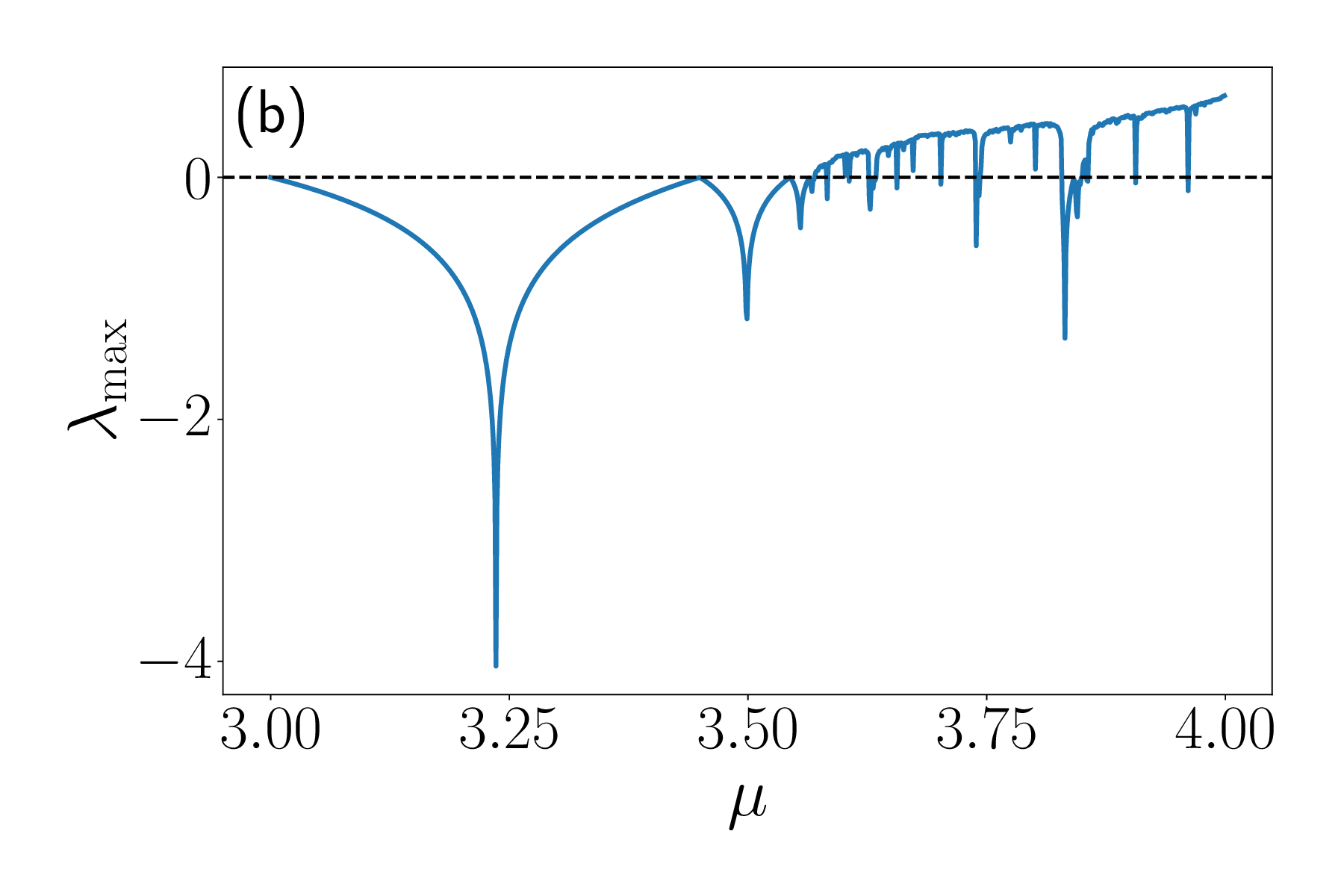}

    \caption{Largest asymptotic Lyapunov exponent as a function of the bifurcation parameter (Eq.~\ref{eq0}). While the asymptotic exponent correctly identifies chaotic and non-chaotic regions, it can not classify the distinct dynamical states, motivating the use of finite-time Lyapunov exponent distributions. (a) the trained RC map computed using $10^5$ time steps, and (b) the logistic map shown for direct comparison. }
    \label{fig2}
\end{figure}
In RC model, an input signal $u(t) \in \mathbb{R}^n$ is embedded into a high-dimensional space of dimension $m \gg n$ through the input weight matrix $W_{in}$. The reservoir states evolve under the combined influence of the projected input and the internal memory of past states. Previous works have shown that incorporating the bifurcation parameter as an additional input allows RC to successfully predict critical phenomena, namely, boundary crises \cite{PhysRevResearch.3.013090} and amplitude death
\cite{PhysRevE.110.034211}. The trained RC map can be expressed as:
\begin{equation}\label{eq2}
\begin{aligned}
    r[i+1] =& (1-\alpha)r[i] + \alpha \hspace{0.1cm} tanh(A r[i] + \\&W_{in}W_{out} r[i] + k_{b} W_b (\varepsilon - \varepsilon_b)).
\end{aligned}
\end{equation}
Here, $\varepsilon$ denotes the bifurcation parameter of the RC map, playing a role analogous to $\mu$ in the logistic map. The matrix $A$ is chosen as the
adjacency matrix of an Erd\H{o}s-R\'enyi random network characterized by a connection
probability $\sigma$ and a spectral radius $\rho$. The matrices $W_{in}$
and $W_b$ are initialized with entries drawn randomly from the intervals $[-b,b]$ and $[-c,c]$, respectively, and remain fixed during training. The RC is trained using chaotic time series generated by the logistic map,
$x_{n+1} = \mu x_n (1 - x_n)$, at parameter values $\mu \in \{3.92, 3.93, 3.94, 3.95\}$. These high dimensional states are then linearly projected back onto the original $n$-dimensional space using the output weight matrix $W_{out}$. The $W_{out}$ matrix is determined via ridge regression by minimizing the prediction error during the training phase. The hyperparameters employed in the simulations are $m = 400$, $\alpha = 0.86$, $b = 2.13$, $c = 1.15$, $\sigma = 0.526$, $\rho = 0.9$, $k_b = 1$, and $\varepsilon_b = 0.3$.

The maximal FTLE distribution for the trained RC map (Eq. \ref{eq2}) is also computed using Eq. \ref{eq1} for different parameter values.

\paragraph{\textbf{Interior crisis in RC:}}
\begin{figure}[t!]
    \centering
    \begin{minipage}{0.49\linewidth}
        \centering
        \includegraphics[width=\linewidth]{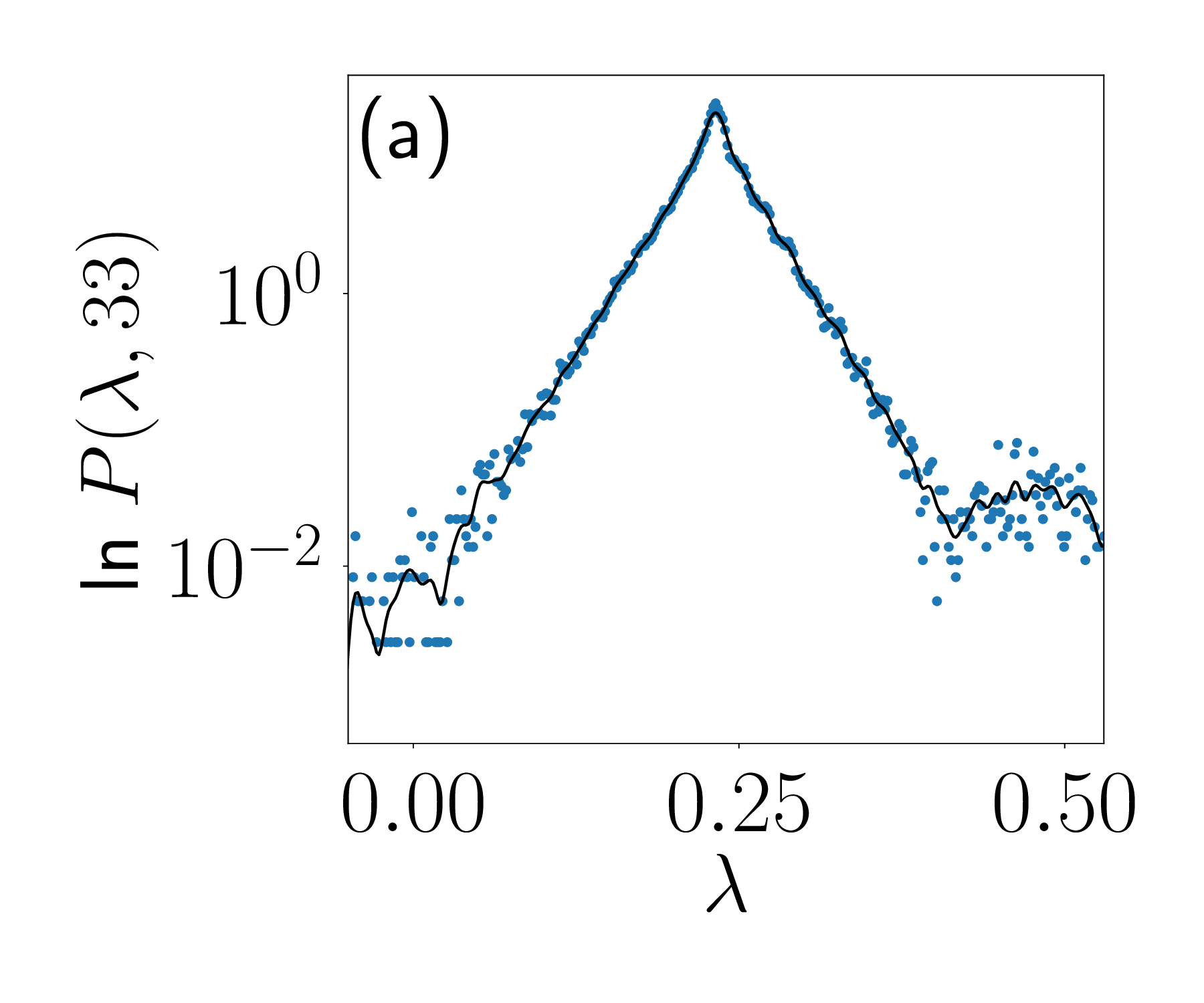}
    \end{minipage}
    \hfill
    \begin{minipage}{0.49\linewidth}
        \centering
        \includegraphics[width=\linewidth]{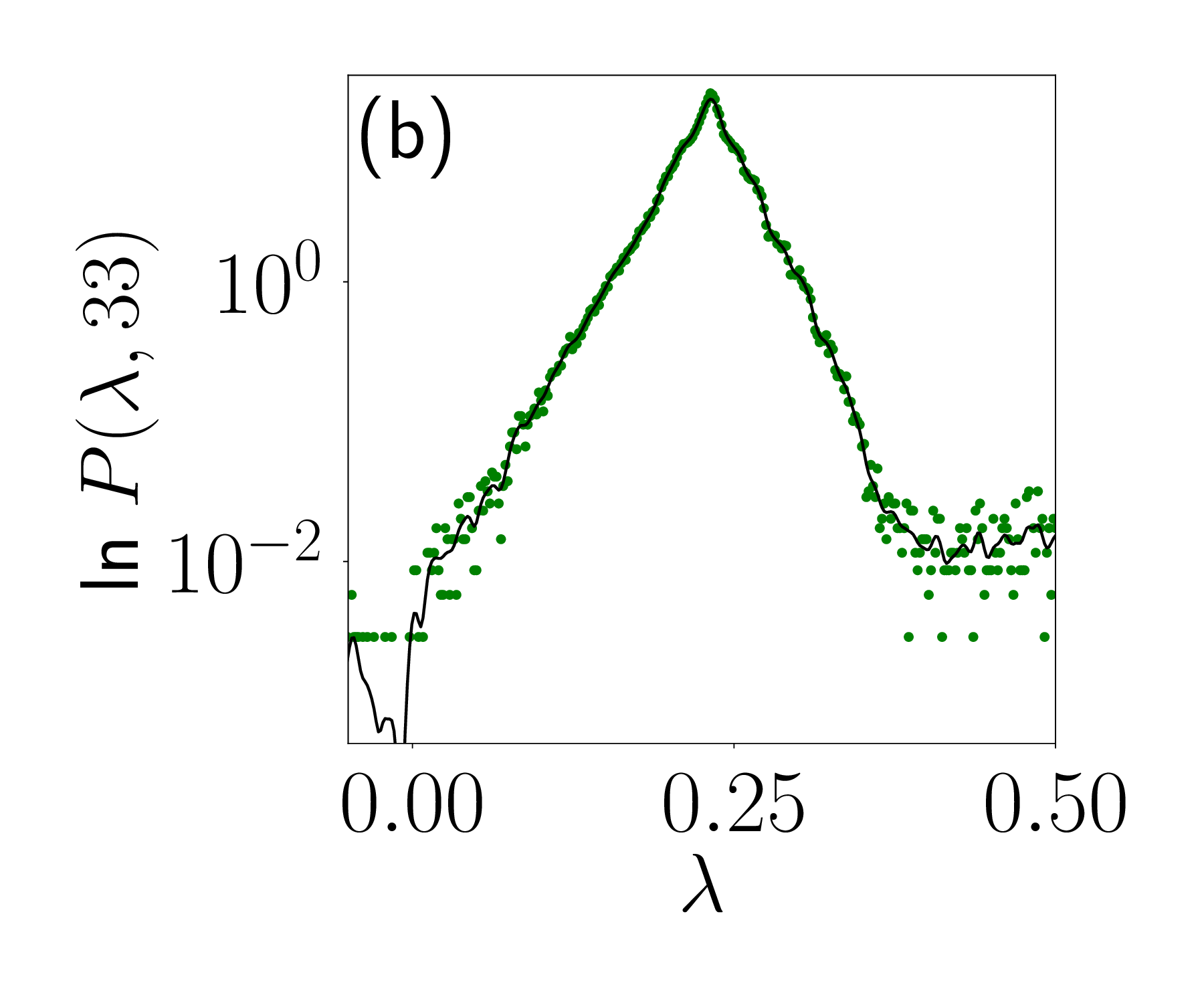}
    \end{minipage}
    \caption{Probability density of finite-time Lyapunov exponents at interior crisis computed from segments of length $N=33$ using $200000$ iterations. 
    (a) Logistic map at $\mu = 3.8568007$; the solid black curve denotes the kernel density estimate for the smooth approximation of the probability density function. 
    (b) FTLE distribution of the trained RC map at $\varepsilon = 3.855865$ closely mimicking the logistic map.}
    \label{fig3}
\end{figure}
We begin by comparing the bifurcation plots of the logistic and the trained RC map. The bifurcation diagram of the trained RC map is constructed by projecting Eq.~\ref{eq2} back into one-dimension using $W_{out}$ matrix. Fig.~\ref{fig1} shows that the RC map closely mimics the logistic map, indicating that periodic windows, chaotic regimes, and other key dynamical features are faithfully captured at the level of asymptotic bifurcation structure. However, such agreement does not ensure consistency in finite-time dynamical properties such as FTLE distributions, nor does it guarantee correct identification of the underlying transition mechanisms, particularly given the high dimensionality of the RC map.

Subsequently, we compute the maximal Lyapunov exponent in the asymptotic limit as a function of the bifurcation parameter for the reservoir map trained on logistic map time series. Despite being trained only with the limited chaotic time series data, the RC map captures the characteristic dips associated with periodic windows with reasonable accuracy (Fig.~\ref{fig2}). The asymptotic maximal Lyapunov exponent provides only coarse information about chaos and fails to capture its rich statistical structure; this motivates a detailed comparison of FTLE distributions between the logistic map and the trained RC map.
\begin{figure}[t!]
    \centering
    \begin{minipage}{0.49\linewidth}
        \centering
        \includegraphics[width=\linewidth]{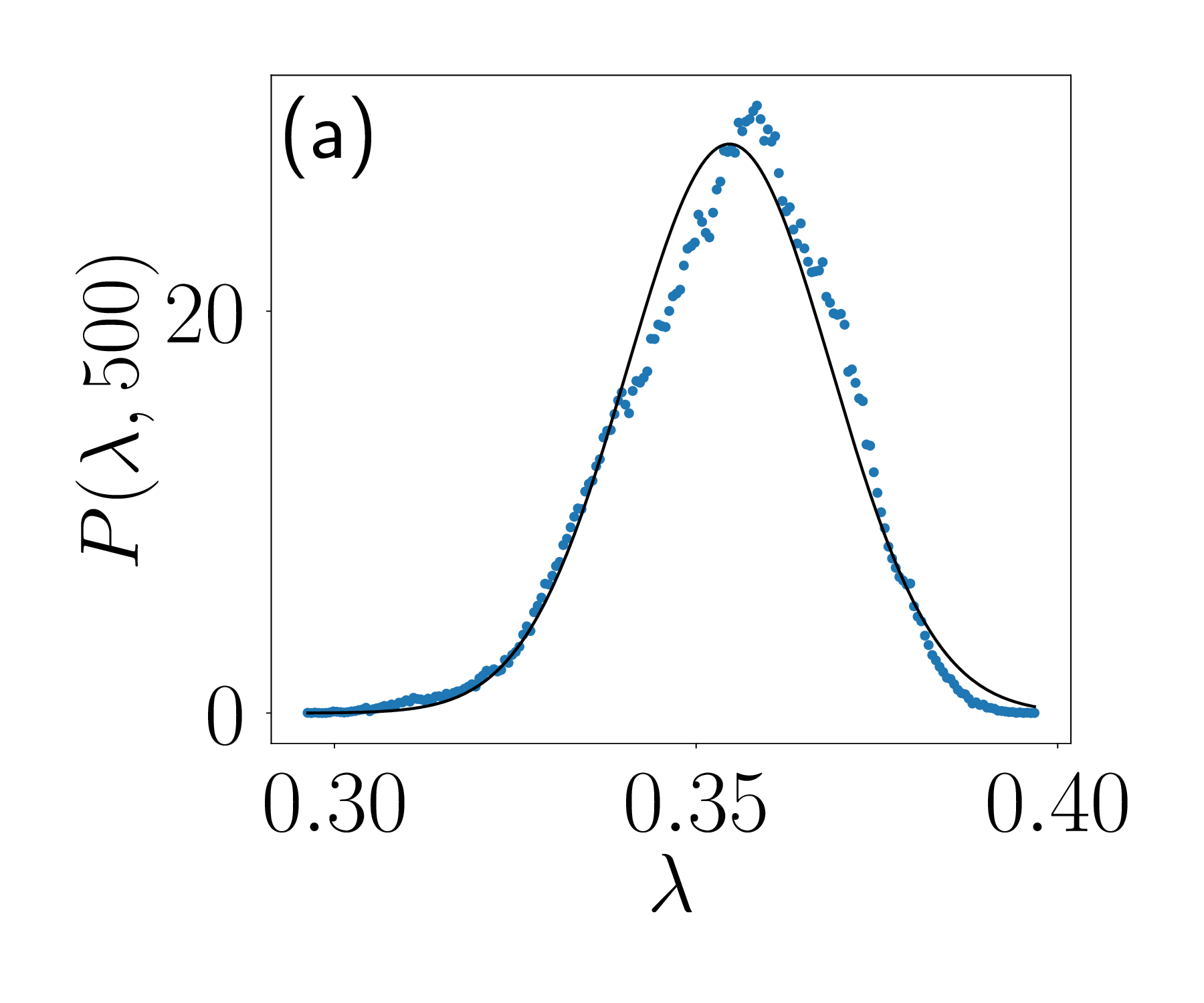}
    \end{minipage}
    \hfill
    \begin{minipage}{0.49\linewidth}
        \centering
        \includegraphics[width=\linewidth]{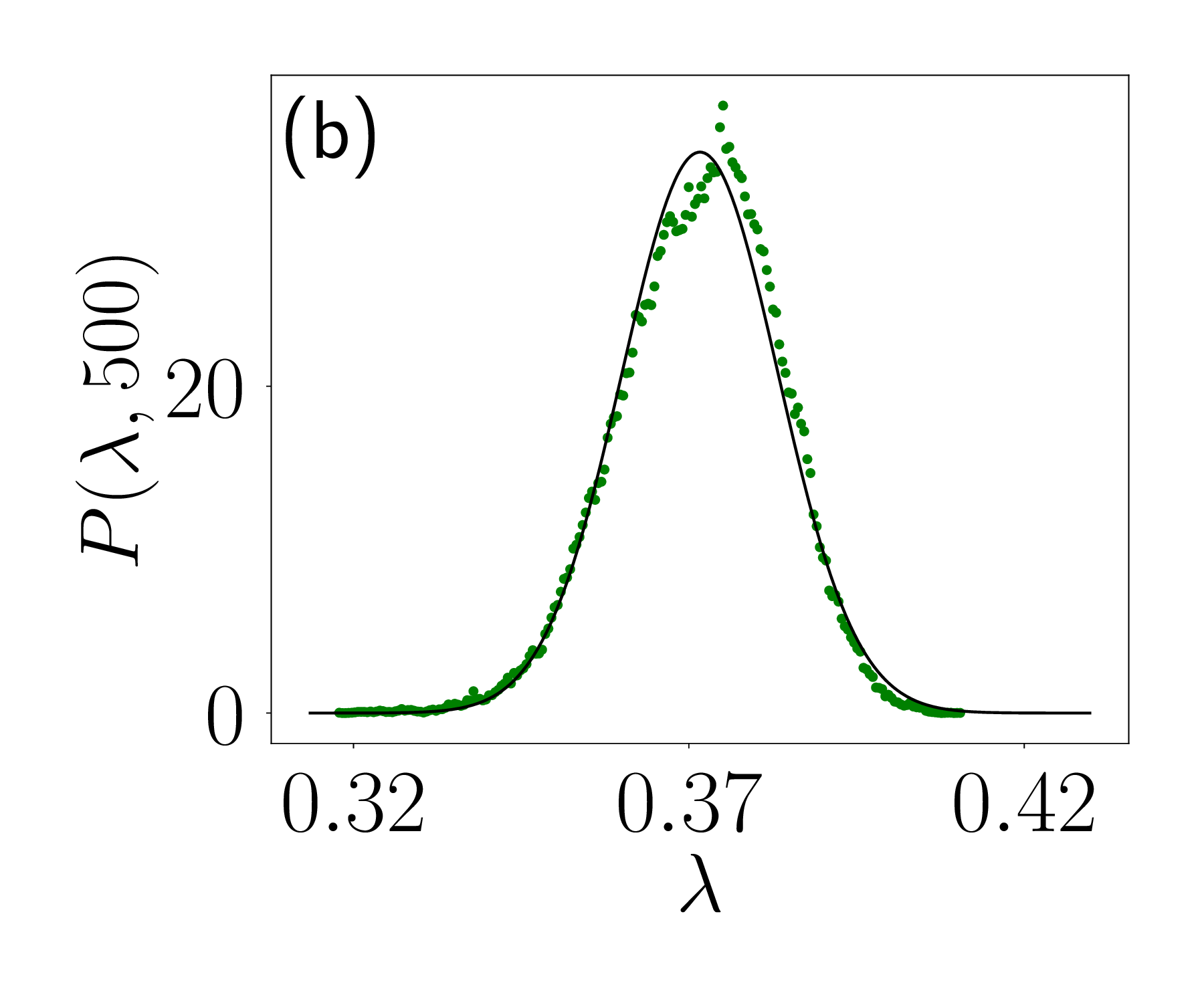}
    \end{minipage}
    \caption{FTLE resemblance in typical chaos regime: 
    Probability density of FTLEs computed from segments of length $N=500$ using $200000$ iterations in the typical chaos regime for    (a) logistic map at $\mu = 3.7$ with the solid black curve denoting the kernel density estimate for the smooth approximation of the probability density function, and
    (b) for trained RC map at $\varepsilon = 3.7$ which matches the gaussian distribution of the input logistic map.}
    \label{fig4}
\end{figure}
We examine the probability distributions of FTLEs in the vicinity of an interior crisis, which is characterized by a sudden expansion of the chaotic attractor \cite{PhysRevLett.48.1507}. In the logistic map, such an interior crisis occurs at $\mu \approx 3.8568007$, where a collision between an unstable period-3 orbit and the chaotic attractor leads to an abrupt increase in attractor size. The FTLE distribution near the interior crisis (Fig.~\ref{fig3} a) can be understood as a superposition of two statistically independent components \cite{PhysRevE.60.2761}: an exponential cusp associated with the pre-crisis chaotic attractor and a Gaussian contribution arising from the newly enlarged attractor. At a parameter value, $\varepsilon = 3.855685$, the trained RC map also exhibits a sudden broadening of the attractor as seen in the bifurcation plot (Fig.~\ref{fig1} b). The trained RC map reproduces the FTLE distribution, closely matching that of the logistic map in the interior crisis regime (Fig.~\ref{fig3} b). Due to the high dimensionality of the reservoir map, it is computationally prohibitive to explicitly identify collisions between unstable periodic orbits and the chaotic attractor. Consequently, the precise mechanism underlying the interior crisis in the RC map cannot be directly verified. Nevertheless, the close correspondence between the FTLE distributions of the logistic map and the trained RC map provides compelling indirect evidence that the reservoir map undergoes an interior crisis driven by a dynamically analogous mechanism, despite the underlying differences in system dimensionality. In this sense, the FTLE distribution serves as an effective proxy for diagnosing the presence of an interior crisis without the need for computationally expensive detection of periodic-orbit collisions. Motivated by this observation, we next examine FTLE distributions across different values of the bifurcation parameter to assess whether the trained reservoir consistently reproduces the statistical signatures associated with distinct dynamical regimes.

\paragraph{\textbf{FTLE distributions in the chaotic regime:}}
In the regime of ``typical" chaos (chaotic regions in $3.57 < \mu < 4$), the Lyapunov exponents behave as random variables due to the rapid decay of correlations, leading FTLEs to obey the central limit theorem \cite{Ott2002}. Consequently, their probability distributions are expected to be Gaussian, with the maximum centered at $\lambda = \Lambda$, where $\Lambda$ denotes the asymptotic Lyapunov exponent in the limit $t \to \infty$. The logistic map at $\mu = 3.7$ indeed exhibits such a Gaussian distribution, which is accurately reproduced by the trained reservoir map at the same parameter value (Fig.~\ref{fig4}(a) and (b)).
\begin{figure}[t!]
    \centering
    
    \begin{tabular}{c@{\hspace{1mm}}c}
        \includegraphics[width=0.48\columnwidth]{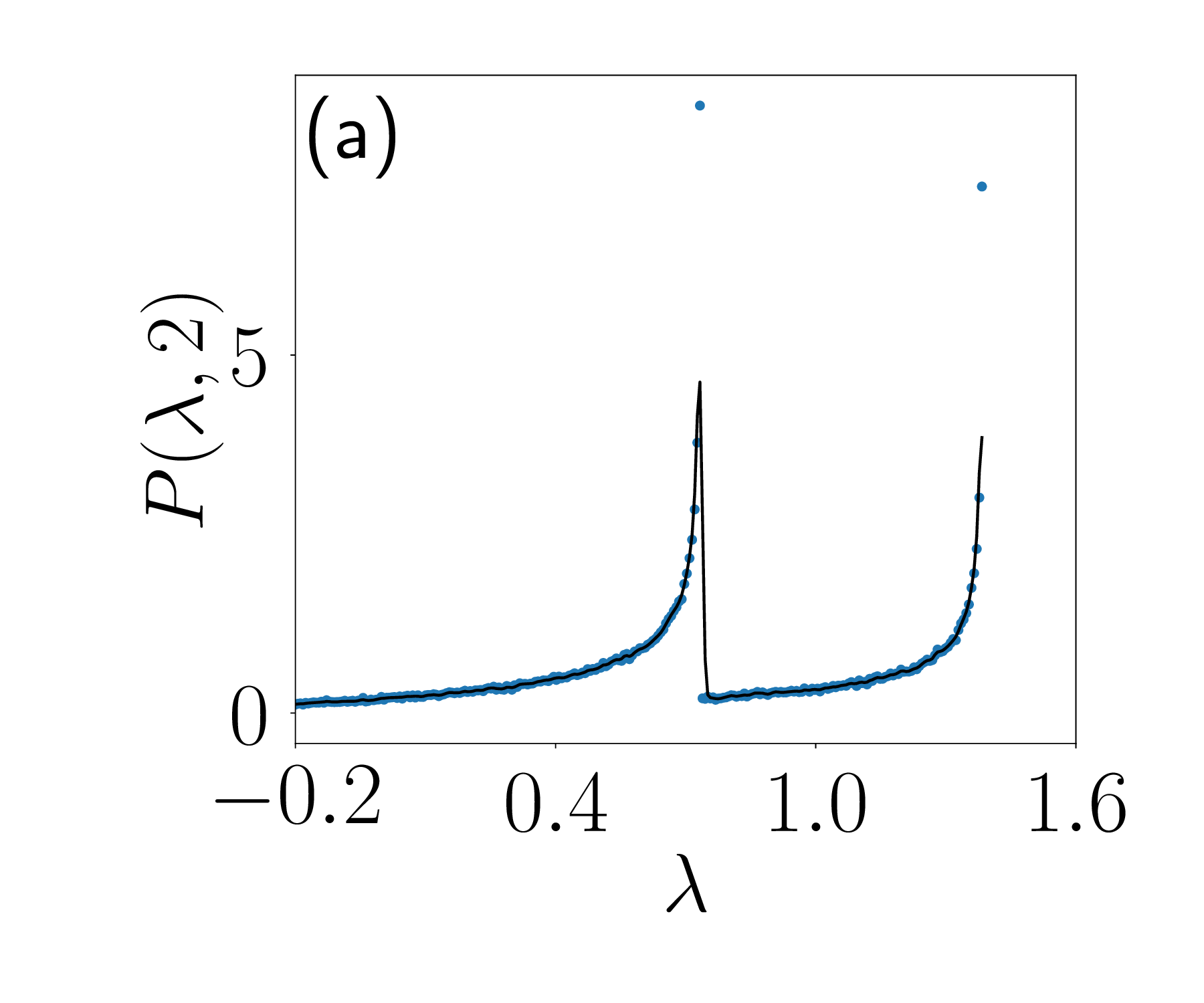} &
        \includegraphics[width=0.48\columnwidth]{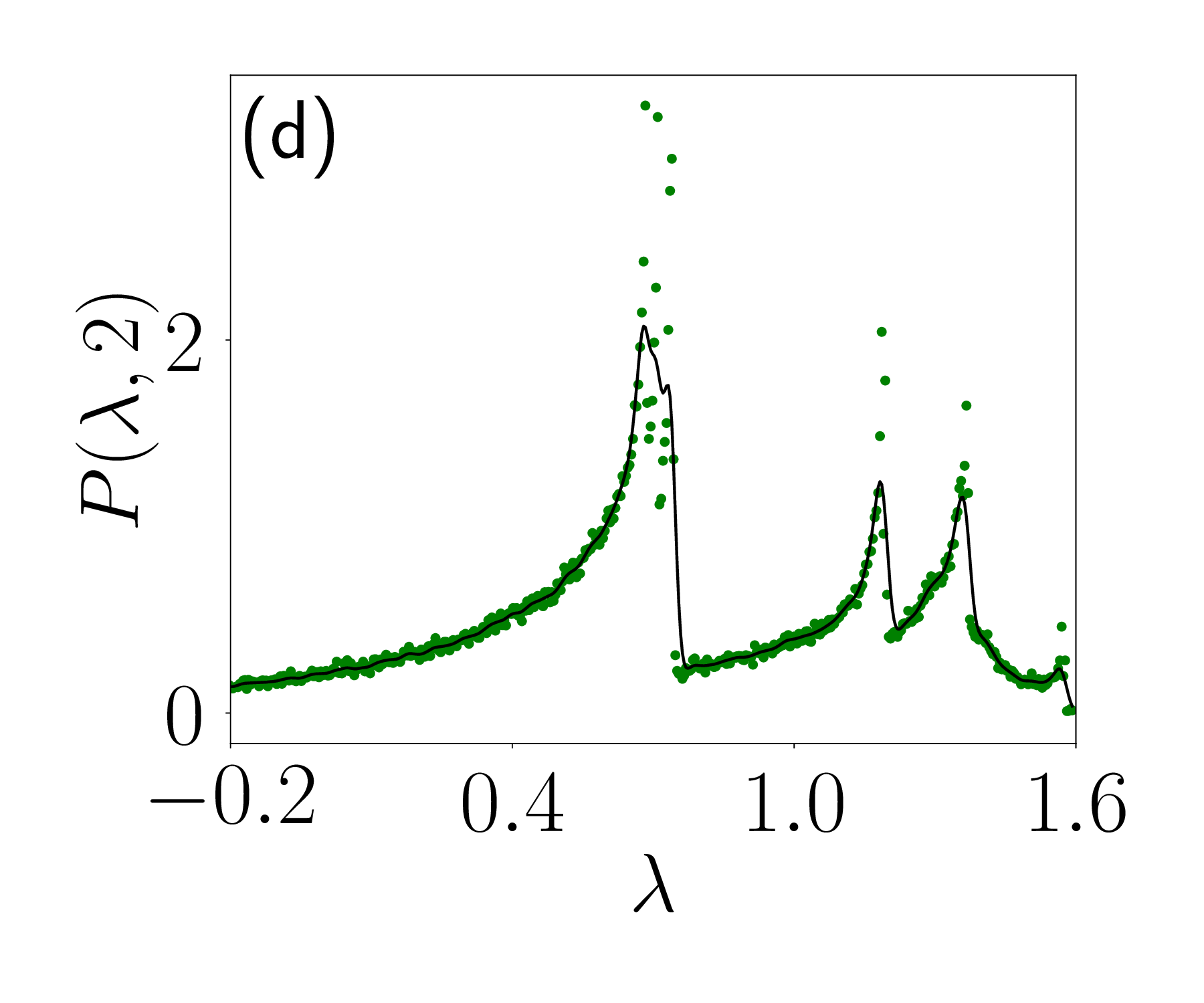} \\[-3mm]
        
        \includegraphics[width=0.48\columnwidth]{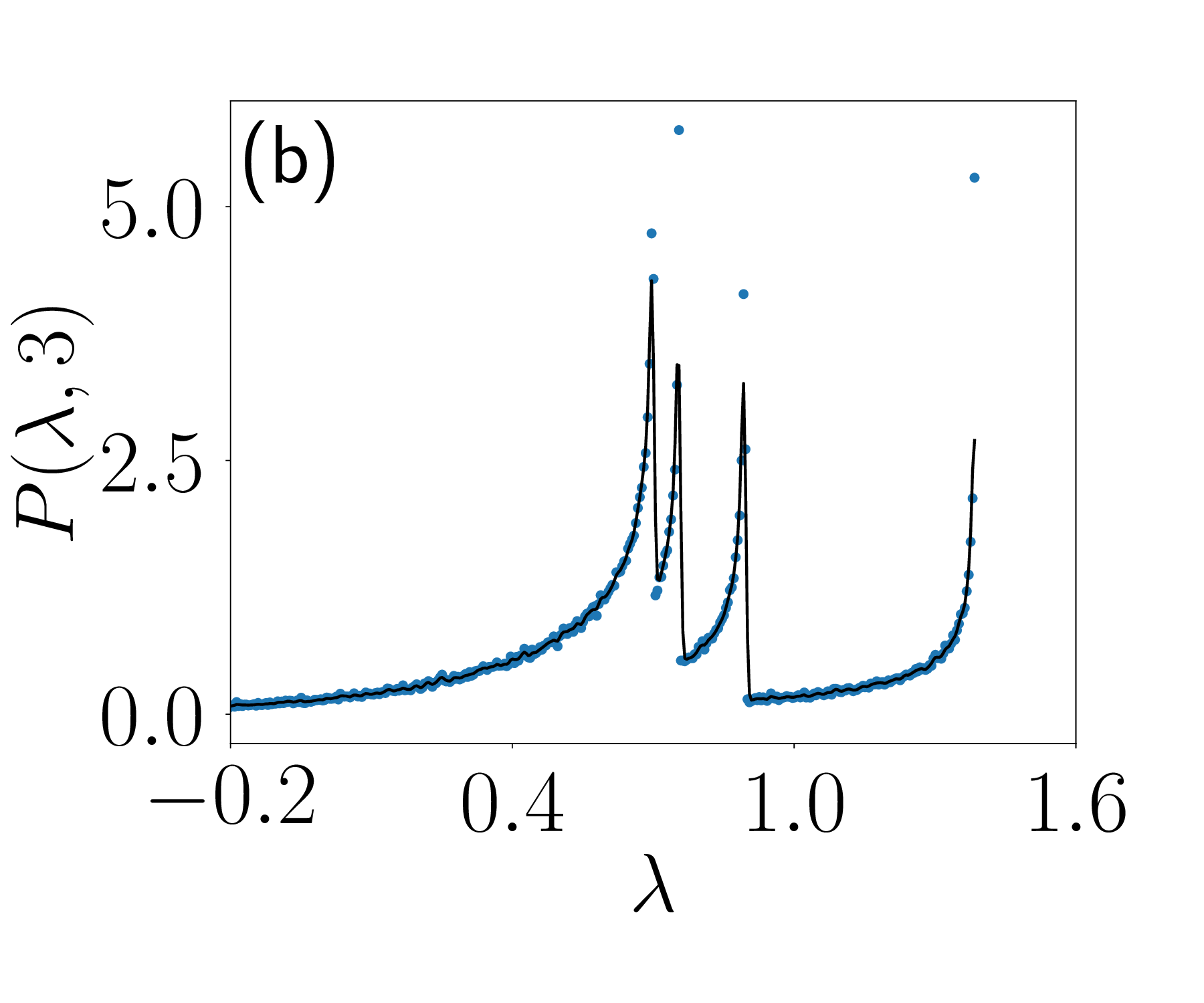} &
        \includegraphics[width=0.48\columnwidth]{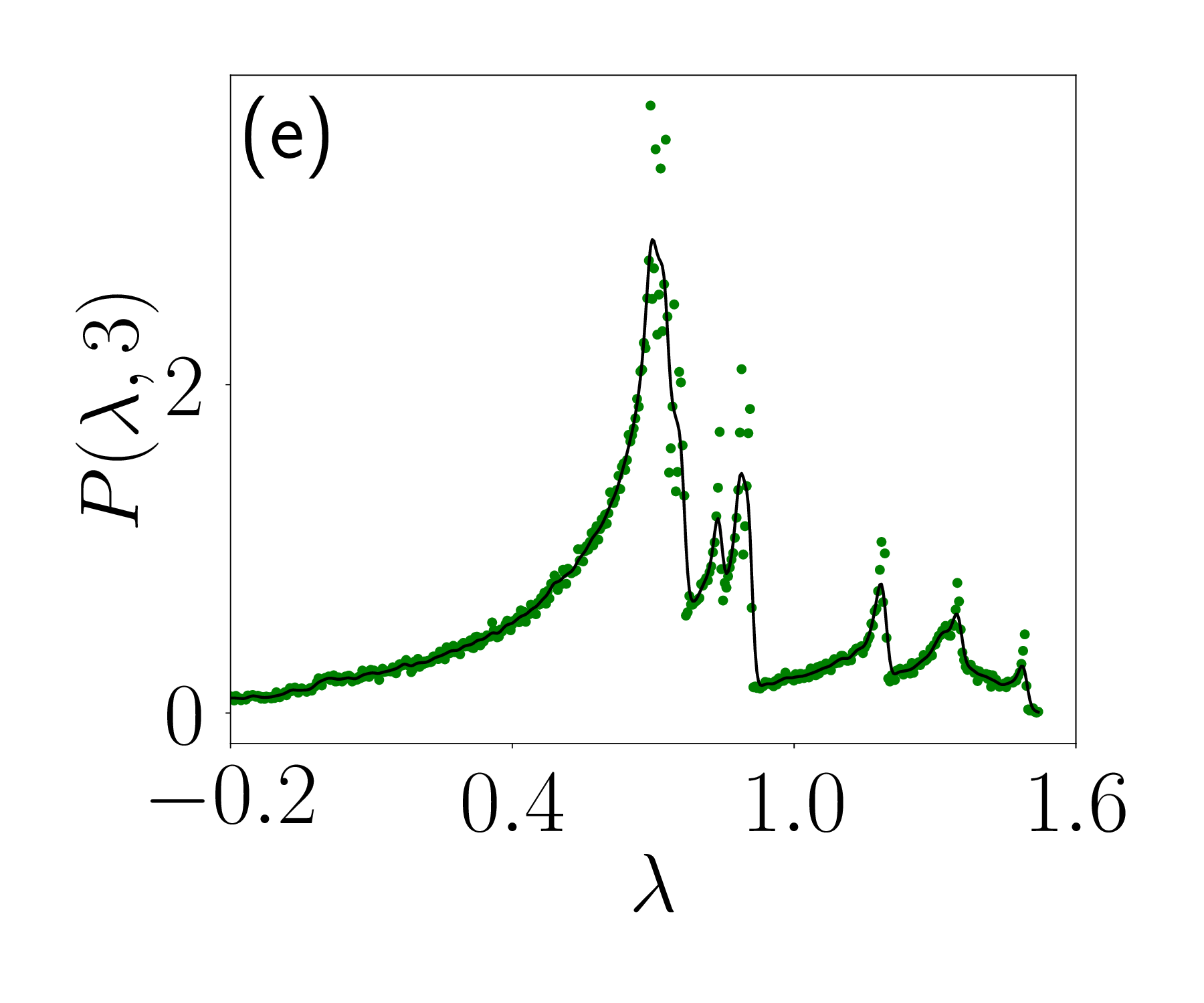} \\[-3mm]
        
        \includegraphics[width=0.48\columnwidth]{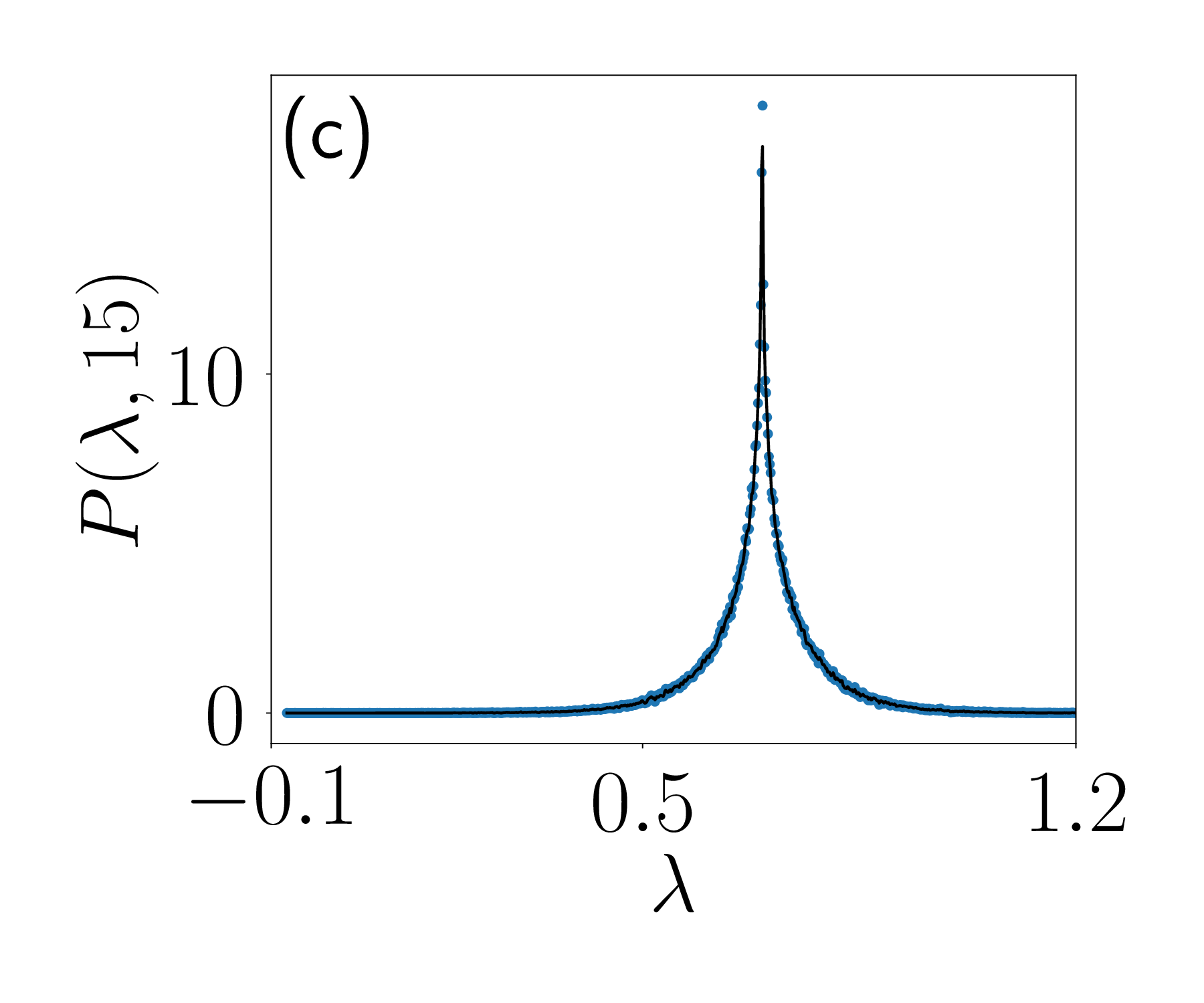} &
        \includegraphics[width=0.48\columnwidth]{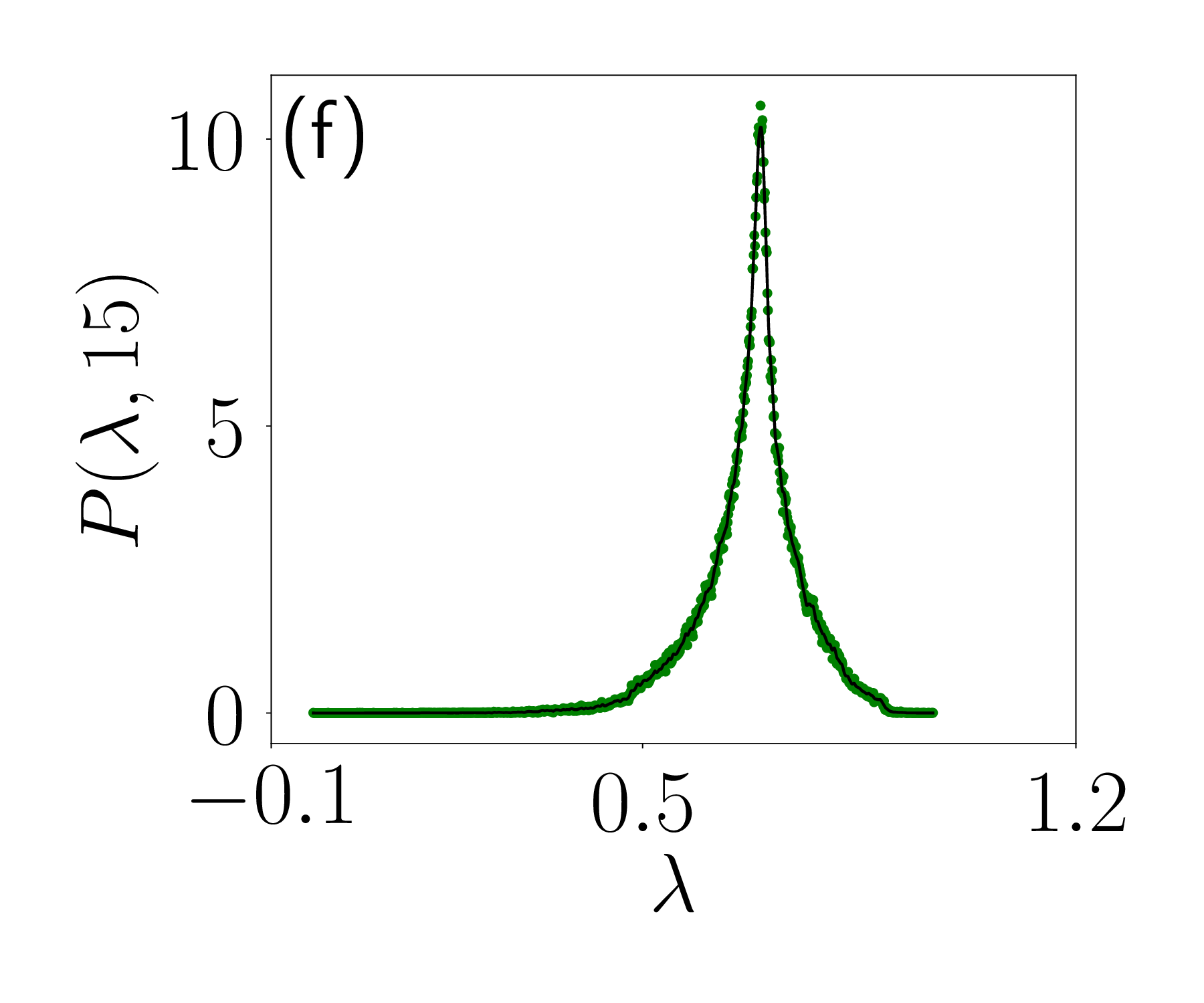}
    \end{tabular}
    \caption{(a) - (c) Probability density estimation of FTLEs at $\mu = 4$ (fully developed chaos) for the logistic map from segments of length a) $N = 2$, b) $N = 3$, c) $N = 15$ showing different number of spikes. (d) - (f) Probability density estimation of FTLEs at $\varepsilon = 3.9985$ for the trained RC map from segments of length d) $N = 2$, e) $N = 3$, f) $N = 15$ showing different number of spikes. The solid black curve denotes the kernel density estimate for the smooth approximation of the probability density function.}
    \label{fig5}
\end{figure}
\begin{figure}[t!]
    \centering
    \includegraphics[scale=0.26]{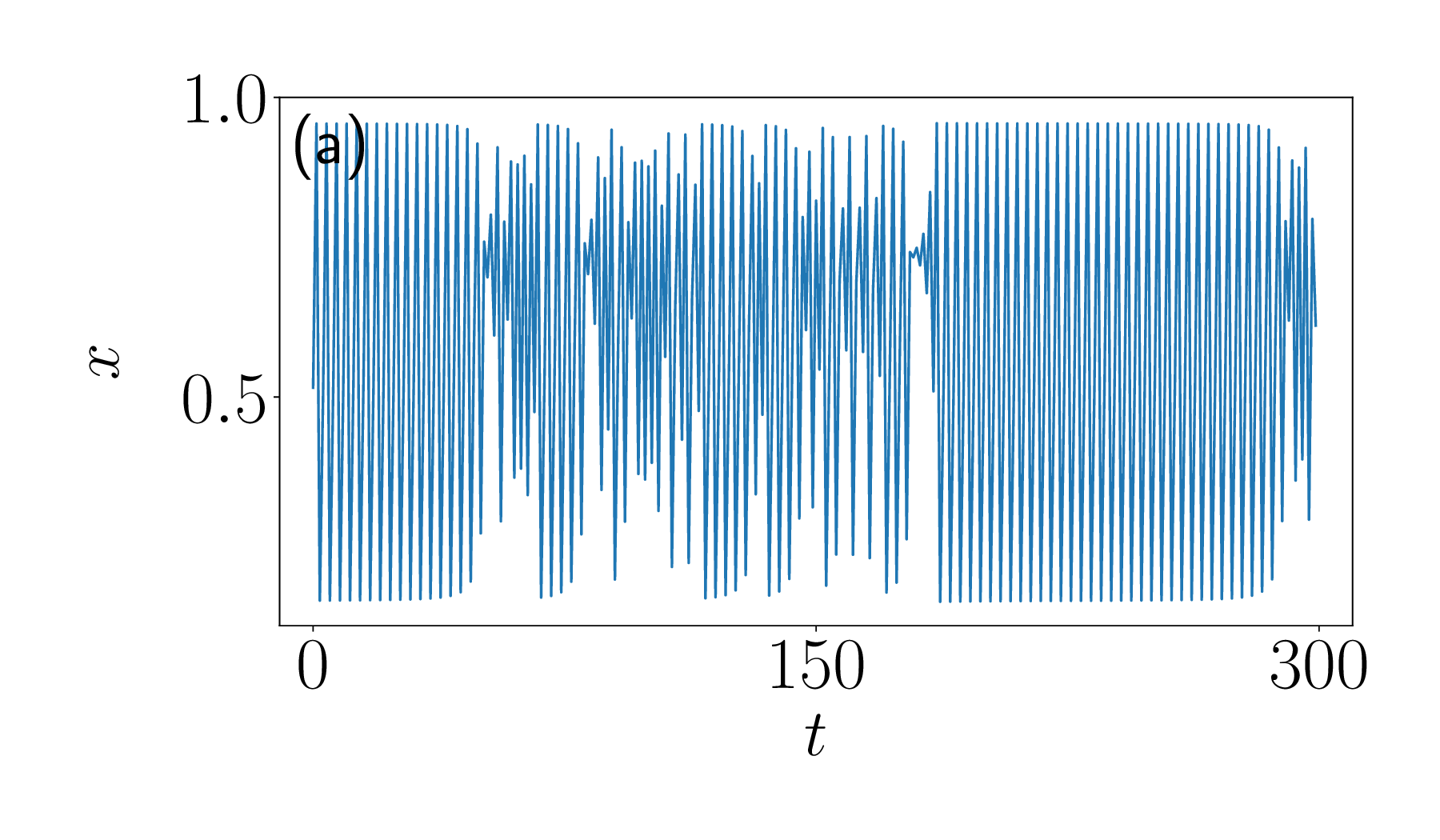}\\
    \vspace{-5mm}
    \includegraphics[scale=0.26]{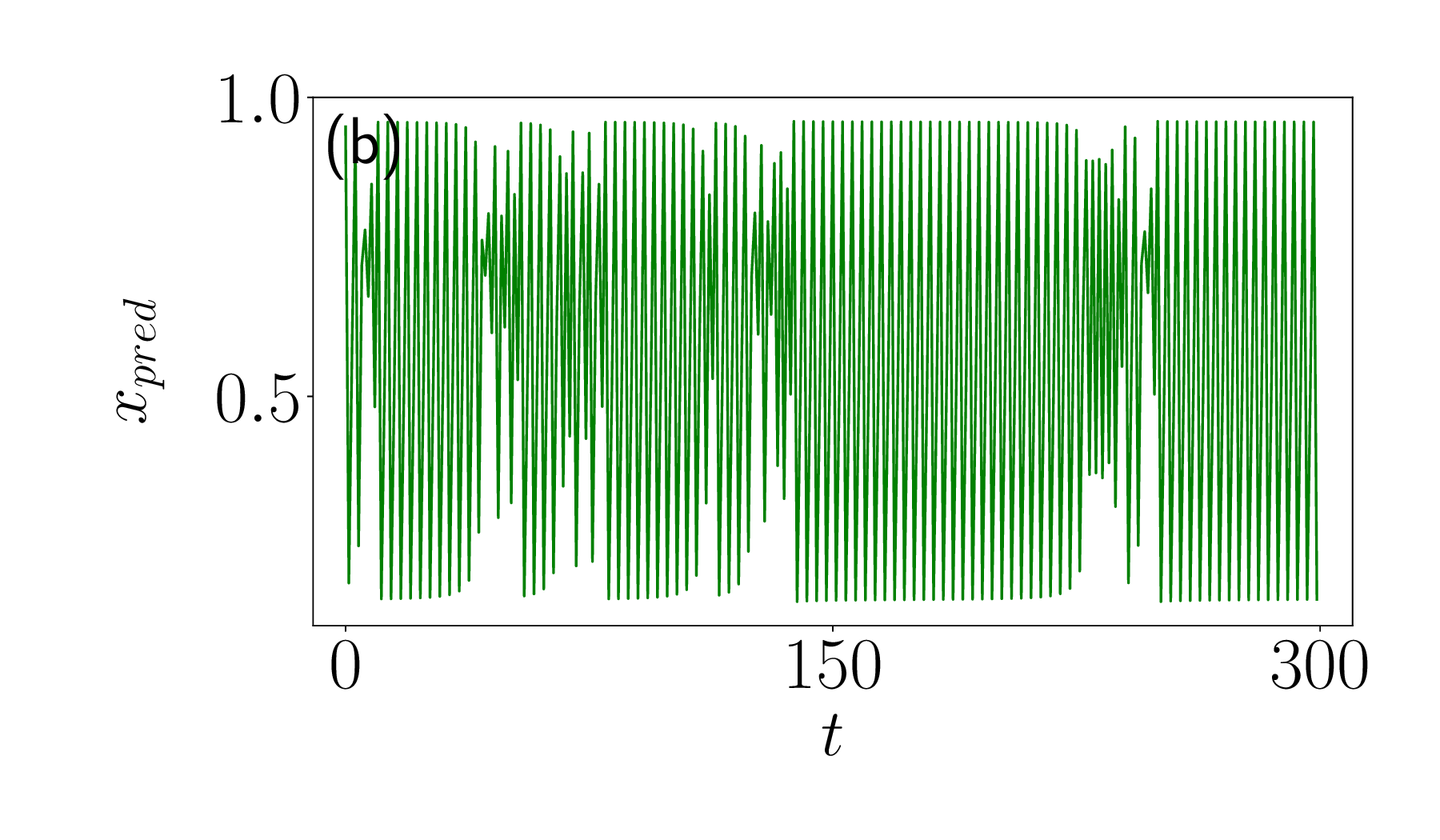}

    \caption{(a) Time series of the logistic map at $\mu = 1 + \sqrt{8} - 10^{-4}$ depicting intermittent behavior due to subduction. (b) Time series of the trained RC map at $\varepsilon = 3.8261$ also manifesting intermittent dynamics.
    }
    \label{fig6}
\end{figure}
The distribution deviates sharply from a Gaussian shape at the Ulam point ($\mu = 4$) also known as ``fully developed chaos", resulting in a cusp at $\lambda = \Lambda$, where $\Lambda = ln 2$ is the Lyapunov exponent in the $t \to \infty$ limit. The analytical origin of the singularities was elucidated in \cite{PhysRevE.60.2761,PhysRevE.69.016207}. For a 1-D map $x_{n+1} = f(x_n)$, FTLE for $n$ steps can be computed as:
\begin{equation}
    \lambda_n = \frac{1}{n} \sum_{j = 0}^{n-1} ln |f'(x_j)| = \frac{1}{n} ln [G_n(x)],  \nonumber
\end{equation}
where $x_j = f^j (x_0)$, the $j^{th}$ iterate of the initial condition $x_0$. The probability distribution of FTLEs has been shown to obey the following equation \cite{PhysRevE.69.016207}:
\begin{equation}\label{eq3}
    P(\lambda, n) = n e^{n\lambda_n} \sum_{roots} \frac{\rho(x)}{|G_n^{'} (x)|}, \nonumber
\end{equation}
where $\rho(x)$ denotes the invariant density of the 1-D map and the sum is performed over all the roots of the polynomial $G_n(x) - e^{n\lambda_n}$. For the logistic map, where $f(x) = 4x(1-x)$, the function $G_n(x)$ is a polynomial of degree $2^n - 1$. For sufficiently large values of $\lambda$, pairs of roots of $G_n(x)$ move off the real axis, resulting in vanishing derivatives and consequently giving rise to singularities in the FTLE distribution. As a result, $P(\lambda, n)$ exhibits $2^{n-1}$ spike like cusps; however, with increasing $n$, these cusps become progressively narrower, and due to the finite resolution imposed by histogram binning, not all spikes can be resolved for large $n$. As illustrated in Fig.~\ref{fig5}(a)-(c), the logistic map displays the theoretically predicted number of spikes for $n = 2$ and $n = 3$, while for $n = 15$ only a single cusp at $\lambda = \Lambda$ remains visible.

We next investigate whether an analogous structure emerges in the trained RC map, which is nontrivial since, in this case, $G_n(x)$ cannot be represented as a polynomial and the origin of singularities is therefore not immediately evident. Remarkably, the trained RC map also develops a cusp at $\lambda = \Lambda$ (Fig.~\ref{fig5}(f)). Owing to the high-dimensional nature of the RC dynamics, the functional form of $G_n(x)$ differs from that of the map, and consequently the number of spikes expected at smaller values of $n$ does not coincide with the one-dimensional case, as observed in Fig.~\ref{fig5}(d)-(e). This further highlights that agreement at the level of coarse dynamical features does not trivially extend to finer statistical properties.
\begin{figure}[t!]
    \centering
    \begin{tabular}{cc}
        \includegraphics[width=0.48\columnwidth]{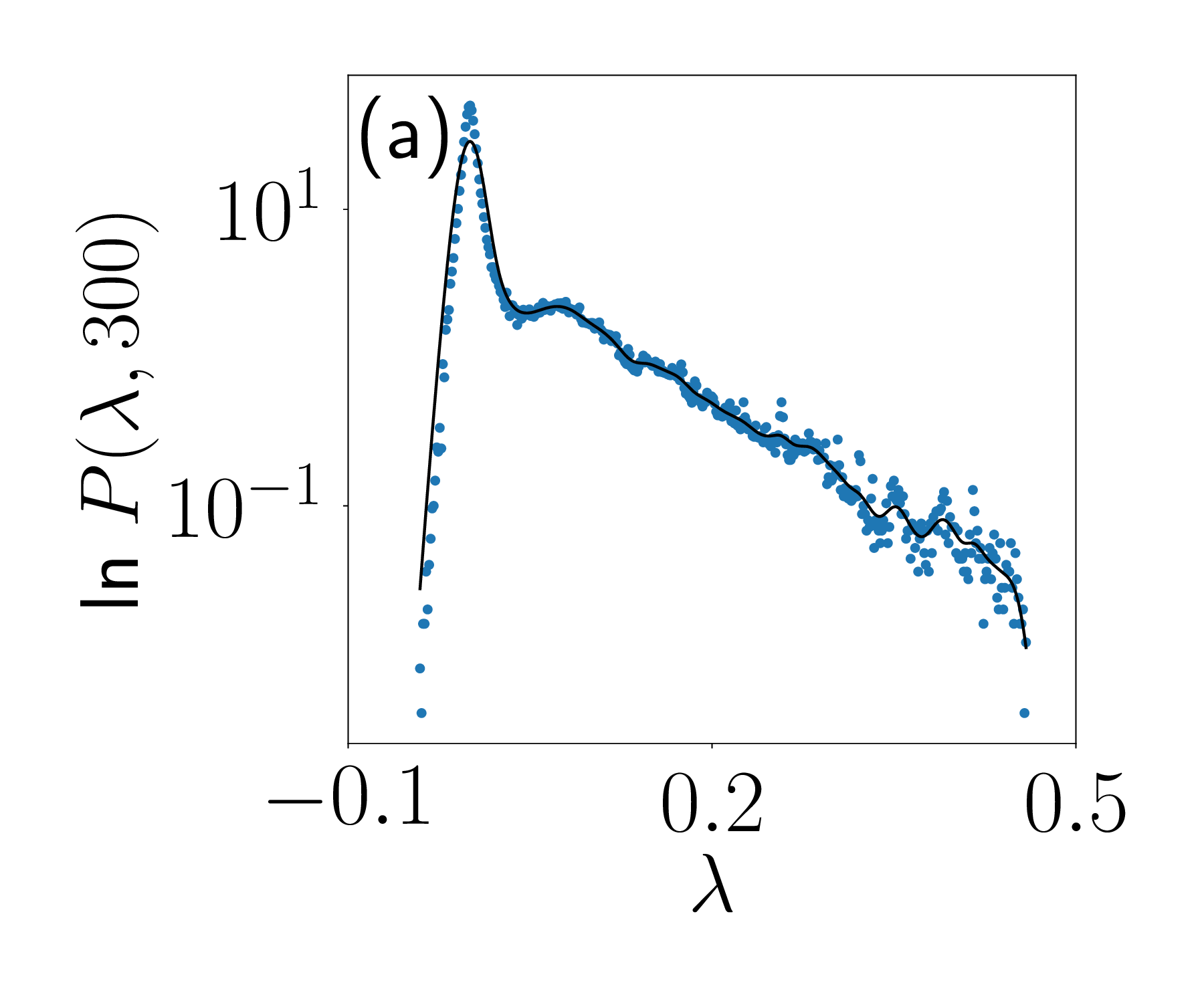} &
        \includegraphics[width=0.48\columnwidth]{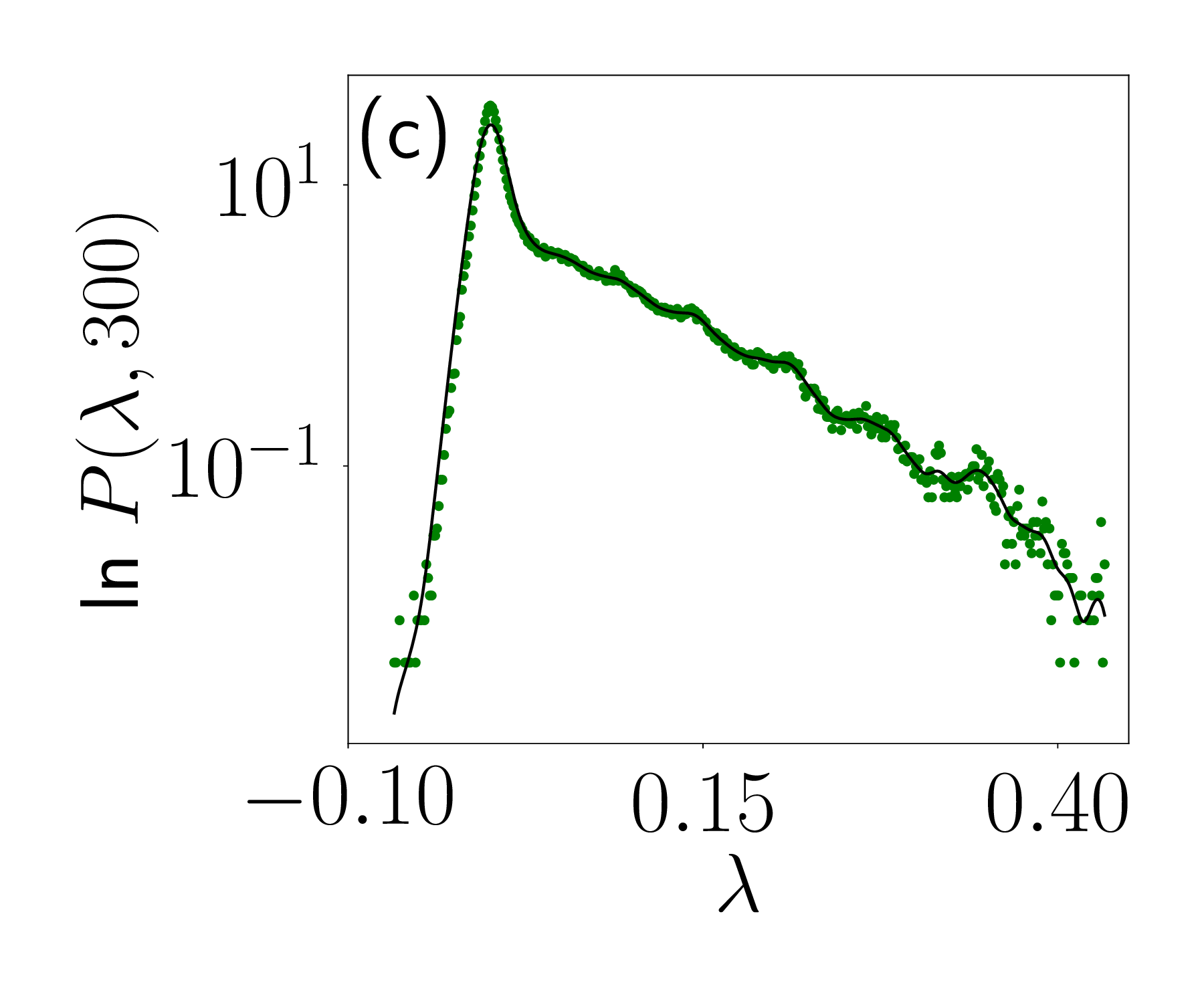} \\[-2mm]
        \includegraphics[width=0.48\columnwidth]{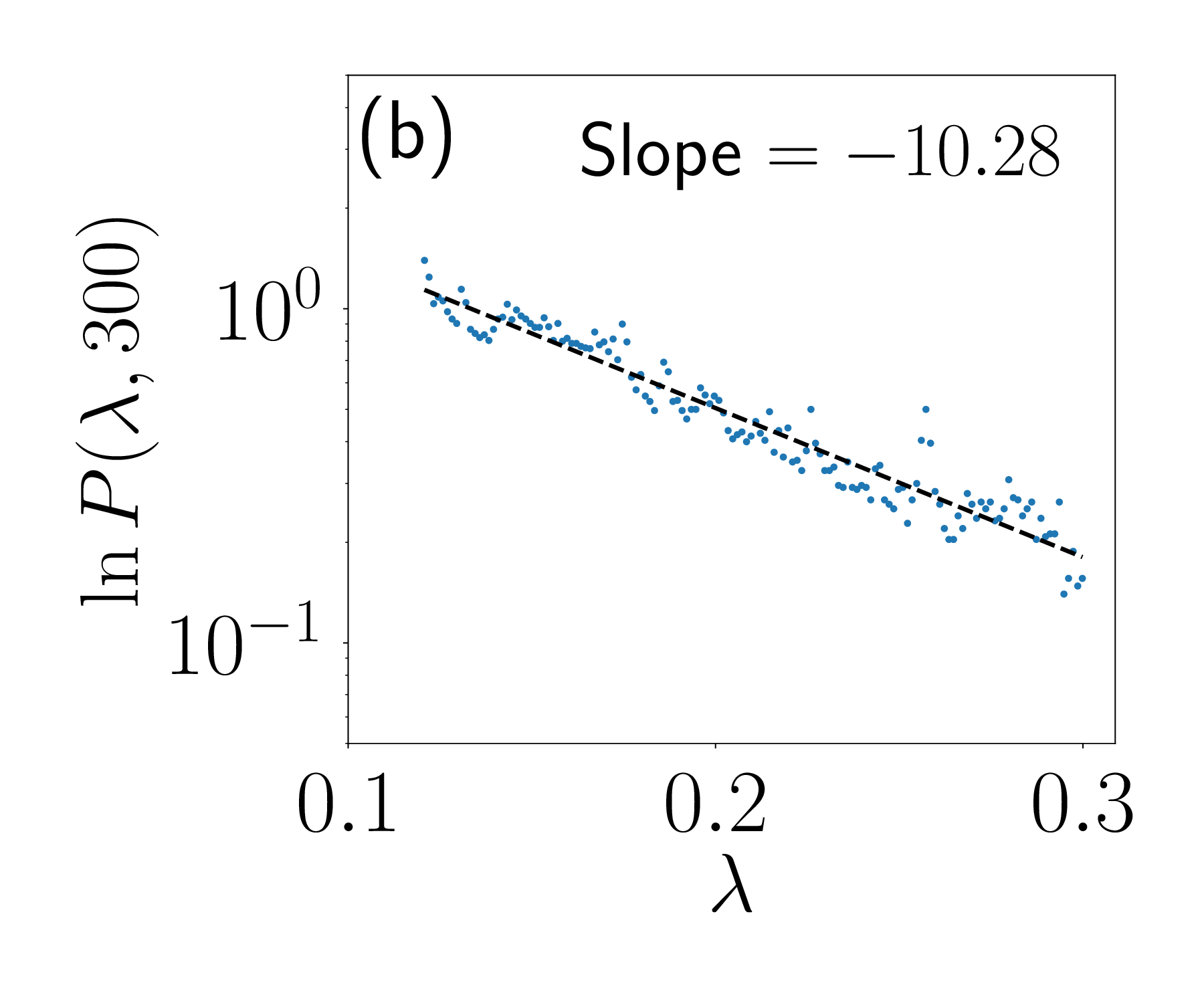} &
        \includegraphics[width=0.48\columnwidth]{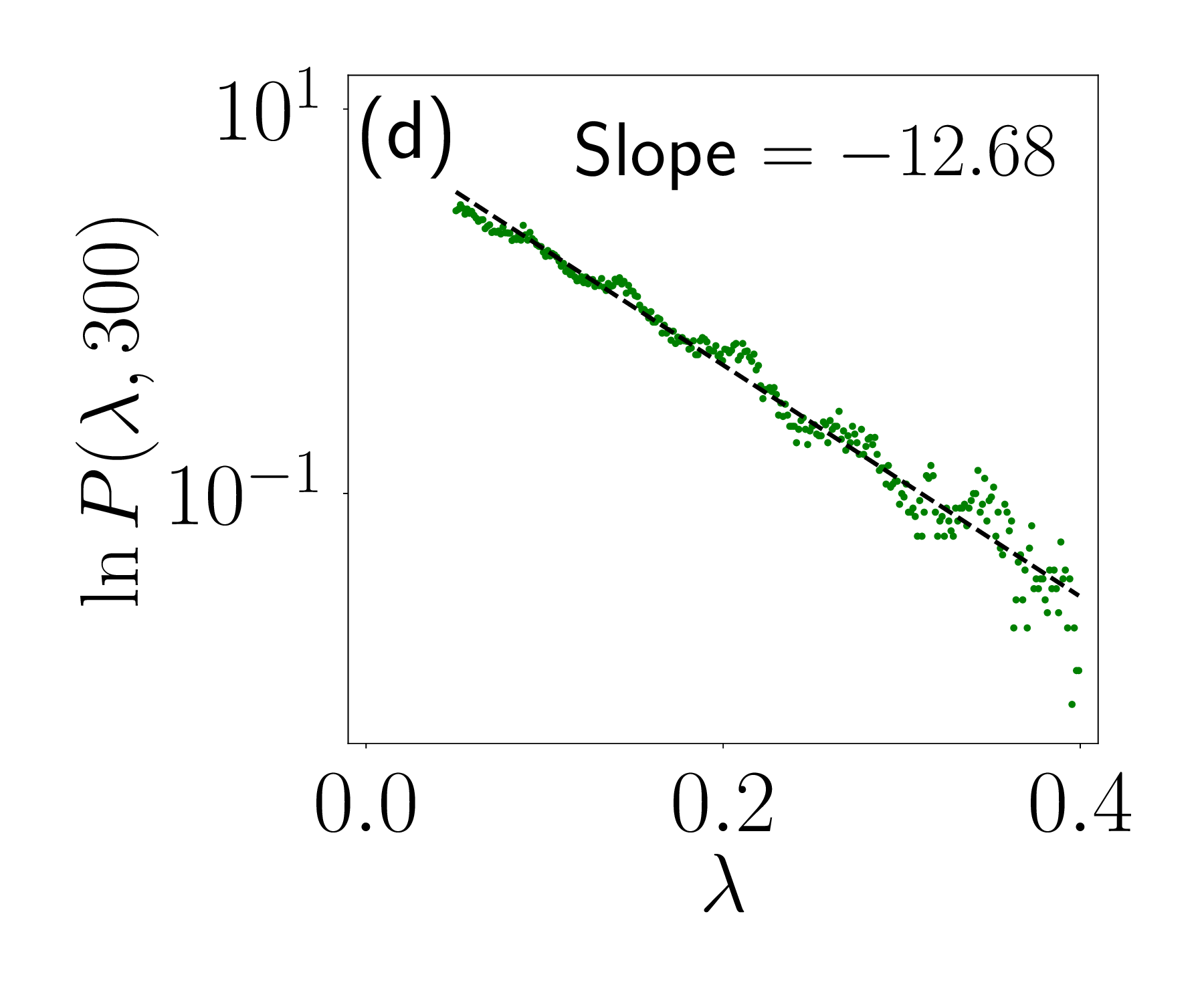} \\
        
    \end{tabular}
    \caption{(a) Probability density estimation of FTLEs at $\mu = 1 + \sqrt{8} - 10^{-4}$ for the logistic map for subduction from segments of length $N = 300$. (b) Straight line fit to the exponential tail to compute the scaling exponent for logistic map. (c) Probability density estimation of FTLEs at $\varepsilon = 3.82625$ for the trained RC map at subduction from segments of length N = 300. (d) Straight line fit to the exponential tail to compute the scaling exponent for RC map.}
    \label{fig7}
\end{figure}

\paragraph{\textbf{FTLE distributions in intermittent regime:}}
The replacement of a chaotic attractor with a non-chaotic attractor without destroying its basin of attraction is termed as subduction \cite{PhysRevLett.48.1507}. It happens in the logistic map at $\mu = 1+ \sqrt{8}$, where the chaotic attractor is replaced by a stable period-3 orbit. A defining dynamical signature of this process is the emergence of intermittency \cite{RevModPhys.53.643} in the time series (Fig.~\ref{fig6} a). Consistent with this behavior, the bifurcation diagram of the trained RC map (Fig.~\ref{fig1} b) reveals the appearance of a period-3 orbit replacing the chaotic attractor at $\varepsilon \approx 3.82625$, and the corresponding time series exhibits clear intermittent dynamics (Fig.~\ref{fig6} b).

The characteristic density of FTLEs in this case appears to be a combination of a normal density and a stretched exponential tail (Fig.~\ref{fig7} a) \cite{PhysRevE.60.2761}. Previous studies have demonstrated that the FTLE distributions for the intermittent dynamics exhibit a universal exponential distribution, in contrast to the typically expected normal distribution about their asymptotic values \cite{Laiftle}. Dynamically, intermittency arises from the alternation between two qualitatively distinct phases: laminar intervals associated with the stable period-3 orbit and chaotic bursts. Each of these regimes contributes a Gaussian component to the FTLE distribution, centered at different values of $\lambda$, while the stretched exponential tail provides a continuous interpolation between them. We demonstrate that the trained RC map not only reproduces the exponential tail (Fig. \ref{fig7} c) but also captures its associated scaling exponent with reasonable accuracy (Fig. \ref{fig7} d), highlighting its ability to faithfully emulate the underlying intermittent dynamics.

\paragraph{\textbf{Conclusion:}}
In this work, we show that the statistical distributions of maximal finite-time Lyapunov exponents provide a sensitive framework for probing transition mechanisms in trained reservoir maps across multiple chaotic regimes. While the high dimensionality of the RC map makes direct verification of unstable periodic-orbit collisions infeasible, the close agreement between the FTLE distributions of the logistic and reservoir maps near an interior crisis provides strong indirect evidence that the trained RC undergoes the same qualitative transition mechanism.

The trained RC map does more than predict trajectories: it also recovers the characteristic statistical features of each regime, such as Gaussian statistics in typical chaotic dynamics, cusp-like singularities at the threshold of fully developed chaos, and exponential tails with the correct scaling exponent in the intermittent regime. This correspondence indicates agreement in finite-time dynamical behavior, which cannot be deduced from bifurcation diagrams alone due to their inherently asymptotic character. We use logistic map as a prototype system as it is capable of exhibiting a range of diverse transitions, serving as an ideal testbed, along with having exact analytical results at the Ulam point enabling stringent comparison with the RC model. The methodology developed here is general and can be applied to higher-dimensional discrete and continuous systems; however, the present work focuses on establishing and validating the approach in a minimal setting where the full range of dynamical scenarios is well understood.

A straightforward direction for future work is to investigate how these characteristic FTLE distributions depend on the hyperparameters of the reservoir, and whether systematic changes in their structure can be used to distinguish between well-trained and poorly trained models, particularly when extending to higher-dimensional systems.
\\

\paragraph{\textbf{Acknowledgment:}}
SJ gratefully acknowledges SERB Power grant award SPF/2021/000136. DS thanks CSIR fellowship under award no. JRF-NET2024/14347.
\bibliography{references}

\end{document}